\documentclass[referee]{raa}
\usepackage{graphicx,times}
\usepackage{natbib}
\usepackage{amssymb,amsmath}
\usepackage{multirow}
\usepackage{hyperref}
\bibpunct{(}{)}{;}{a}{}{,}

\hypersetup{pdftitle = The title of my PDF, pdfauthor = My name, pdfsubject= The subject, pdfkeywords = keyword1 keyword2 keyword3}
\hypersetup{colorlinks = true, linkcolor = green, anchorcolor = red, citecolor = blue, filecolor = red, pagecolor = red, urlcolor = red}

\begin{document}

   \title{First Digit Distributions of Gamma-Ray Bursts}

 \volnopage{Vol.0 (20xx) No.0, 000--000}      
   \setcounter{page}{1}

   \author{Hou-Yu Lai\inst{1,2} \and Jun-Jie Wei\inst{1,2}
   }

   \institute{Purple Mountain Observatory, Chinese Academy of Sciences, Nanjing 210023, China; {\it jjwei@pmo.ac.cn}\\
	\and School of Astronomy and Space Sciences, University of Science and Technology of China, Hefei 230026, China\\
\vs \no
   {\small Received~~20xx month day; accepted~~20xx~~month day}
}

\abstract{The occurrence of the first significant digits from real world sources is usually not equally distributed,
but is consistent with a logarithmic distribution instead, known as Benford's law. In this work,
we perform a comprehensive investigation on the first digit distributions of the duration, fluence,
and energy flux of gamma-ray bursts (GRBs) for the first time. For a complete GRB sample detected by
the Fermi satellite, we find that
the first digits of the duration and fluence adhere to Benford's law. However, the energy flux shows
a significant departure from this law, which may be due to the fact that a considerable part of
the energy flux measurements are restricted by lack of spectral information. Based on the conventional
duration classification scheme, we also check if the durations and fluences of long and short GRBs
(with duration $T_{90}>2$ s and $T_{90}\leq2$ s, respectively) obey Benford's law. We find that
the fluences of both long and short GRBs still agree with the Benford distribution, but their durations
do not follow Benford's law. Our results hint that the long--short GRB classification scheme
does not directly represent the intrinsic physical classification scheme.
}

   \authorrunning{Lai \& Wei}            
   \titlerunning{First Digit Distributions of GRBs}  
   \maketitle

%
\section{Introduction}           
\label{sect:intro}

People might think that the first significant digits (i.e., 1,2,...,9) of any randomly chosen dataset
tend to be uniformly distributed, but it is not true in the natural world. As early as in 1881, Simon Newcomb
had observed an unanticipated pattern in the first digits of logarithm tables: the number 1 appears
more frequently than 2, 2 than 3, and so on \citep{1881AmJM....4...39N}. More than a half century later,
Frank Benford rediscovered that there is a logarithmic distribution of first digits in numerous data
tables, which is often called the first digit law or Benford's law \citep{1938PAPhS..78..551B}.
This law states that for a given real database, the probability of numbers with the first digit $k$
is expressed as \citep{1938PAPhS..78..551B}
\begin{equation}
P(k)=\log_{10}{\left(1+\frac{1}{k}\right)},\quad  k=1,2,...,9 \;.
\label{eq:Benford'law}
\end{equation}

Empirically, Benford's law has been proven in various research fields, including geography
(e.g., the lengths of rivers and the areas of lakes; \citealt{1938PAPhS..78..551B}), finance (e.g.,
the stock market indices; \citealt{Ley1996OnTP,RePEc:eee:empfin:v:5:y:1998:i:3:p:263-279}), biology
(e.g., pre-vaccination measles incidence data, absolute values from human magnetoencephalography
recordings, and gene data lengths of bacteria; \citealt{Cceres2008ContactoC}), seismology (e.g., the recurrence times of
seismic events; \citealt{Sottili2012BenfordsLI}), statistical and nuclear physics (e.g., physical
constants and distributions \citep{1991AmJPh..59..952B,2010PhyA..389.3109S}, half-lives of unstable
nuclei \citep{1993EJPh...14...59B,2008EPJA...38..251N,2009CoTPh..51..713N}, widths of hadrons
\citep{2009MPLA...24.3275S}, and the lepton branching fractions \citep{2018PhyA..506..919D}), etc.
In practice, this peculiar law has been effectively used to distinguish and diagnose frauds in taxing and
accounting \citep{Nigrini1996,Nigrini1997TheUO,doi:10.1081/SAC-120028442}, and to minimize storage
space and speed up calculation in computer science \citep{Barlow1996,Schatte1988OnMD,doi:10.1080/00029890.2007.11920449}.
Theoretically, this law has been well explained by using a central-limit-like theorem for
first digits \citep{5c879bb8-d380-3974-9772-f812aff3bbd5} and a simple Markov process
\citep{2021AmJPh..89..851B}. In mathematics, Benford's law is scale invariant \citep{Berger2008},
which indicates that it is independent of any particular choice of units \citep{Pinkham1961OnTD}.

In astronomy, Benford's law also has extensive application in all sorts of astrophysical datasets,
such as light curves of variable stars and other X-ray sources \citep{2006IJMPC..17.1597M},
pulsar properties \citep{2010APh....33..255S}, distances of galaxies and stars \citep{2014JApA...35..639A},
exoplanetary and asteroid data \citep{2017JApA...38....7S,2021NewA...8901654M}, GAIA DR2 parallaxes
\citep{2020A&A...642A.205D}, and so on. Nevertheless, there are also some types of data, e.g.,
pulsar and fast radio burst dispersion measures, which do not obey Benford's law \citep{2023APh...14402761M}.

In this work, we investigate the first digit distributions of the duration, fluence, and energy flux of
gamma-ray bursts (GRBs), and check if these first digits conform to Benford's law for the first time.
GRBs are flashes of high-energy radiation originating from energetic explosions in the Universe. According to
their duration time $T_{90}$ (the time interval observed to contain 90\% of the prompt emission), GRBs can be
classified into long GRBs ($T_{90}>2$ s) and short GRBs ($T_{90}\leq2$ s) \citep{1993ApJ...413L.101K}.
Generally, long GRBs are suggested to be powered by the core collapses of massive stars \citep{1998ApJ...494L..45P,2006ARA&A..44..507W},
and short GRBs by the mergers of binary compact objects \citep{1989Natur.340..126E,1992ApJ...395L..83N}.

\section{Observational data and statistical results}
\label{sec:data}
\subsection{Dataset}
We download the durations $T_{90}$ (in units of s), fluences $F$ (in units of erg $\rm cm^{-2}$),
and energy fluxes $P_{\gamma}$ (in units of erg $\rm cm^{-2}$ $\rm s^{-1}$) in the 10--1000 keV
energy range from the online catalog of GRBs observed by Fermi's Gamma-ray Burst Monitor (Fermi-GBM)
\citep{2014ApJS..211...12G,2014ApJS..211...13V,2016ApJS..223...28N,2020ApJ...893...46V}.\footnote{\url{https://heasarc.gsfc.nasa.gov/W3Browse/fermi/fermigbrst.html}}
The Fermi-GBM burst catalog comprises of a list of 3665 cosmic GRBs between 2008 July 12 and 2023
December 6. We remove 1 GRB for which no relative data were available. We carry out the first digit
analyses for all the remaining 3664 GRBs with $T_{90}$ and fluence measurements. The energy flux
of each burst in the observer frame is calculated as
\begin{equation}
P_{\gamma}=p_{64}\times\frac{\int_{E_{\rm min}}^{E_{\rm max}}E\times N(E){\rm d}E}{\int_{E_{\rm min}}^{E_{\rm max}}N(E){\rm d}E}\;,
\end{equation}
where $p_{64}$ is the peak flux on the 64 ms timescale (in units of photon $\rm cm^{-2}$ $\rm s^{-1}$),
the spectral model $N(E)$ is the Band function \citep{1993ApJ...413..281B}, $E_{\rm min}$ and $E_{\rm max}$
are 10 keV and 1000 keV, respectively. It is obvious that the spectral parameters
are required to calculate the energy flux. But not every burst has the spectral information.
There are only 2298 GRBs with energy flux measurements in the catalog.

\subsection{Results}
\begin{figure*}
\vskip-0.2in
   \centering
  \includegraphics[width=70mm]{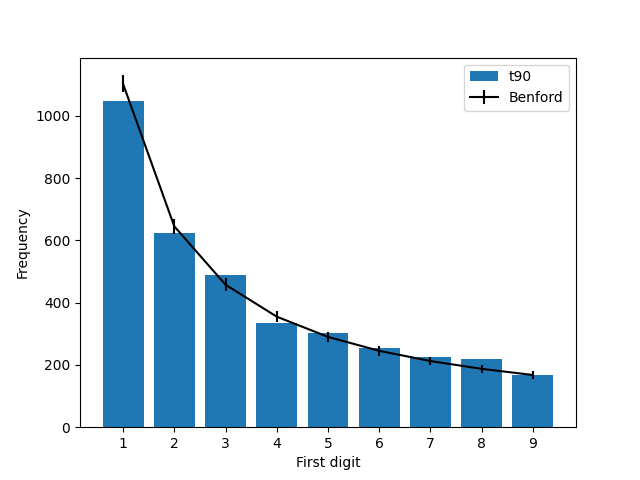}
  \includegraphics[width=70mm]{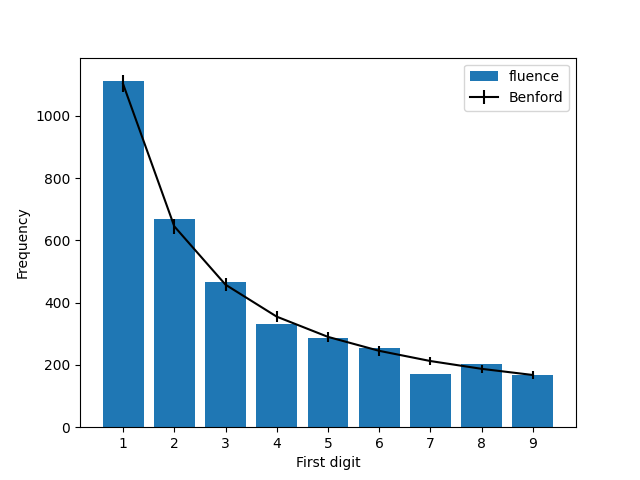}
  \vskip-0.1in
\caption{Distributions of the first digit of the duration $T_{90}$ and fluence of all $3664$ GRBs from the Fermi-GBM burst catalog.
The theoretical predictions from Benford's law (solid lines) along with associated binomial errors are also shown for comparisons.}
\label{fig1}
\end{figure*}

The first digit distributions of the duration and fluence for the complete GRB sample are presented in Figure~\ref{fig1}
and Table~\ref{tab1}. As described above, there are totally $N_{\rm tot}=3664$ available GRBs. The expected number
according to Benford's law, $N_{\rm Ben}=N_{\rm tot} P(k)$, along with the root mean square error estimated by
the binomial distribution, $\Delta N=\sqrt{N_{\rm tot}P(k)(1-P(k))}$, are also shown in the figure. From Figure~\ref{fig1},
we can see that the observed distributions are well consistent with the theoretical predictions from Benford's law.
In order to quantify the goodness of fit of Benford's law, we adopt the Pearson $\chi^2$
\begin{equation}
\chi^2=\sum_{k=1}^{9}\frac{\left[N_{\rm obs}(k)-N_{\rm Ben}(k)\right]^2}{N_{\rm Ben}(k)} \;,
\end{equation}
where $N_{\rm obs}$ and $N_{\rm Ben}$ are the observed number and the expected Benford number for a single digit $k$, respectively.
For the first digit distribution of the duration, we obtain a Pearson $\chi^2$ value of 13.0 for 8 degrees of freedom.
For the fluence distribution, we obtain a $\chi^2$ value of 12.5. These two $\chi^2$ values correspond to $p$-values of
0.11 and 0.13, respectively, which strongly support the null hypothesis that the durations and fluences of the complete
GRB sample follow Benford's law. It is worth emphasizing that the higher the $p$-value, the more likely the null hypothesis.
In our present study, we exclude a null hypothesis if $p<0.05$ (equivalent to under the 95\% confidence level).

\begin{figure*}
\vskip-0.2in
   \centering
  \includegraphics[width=70mm]{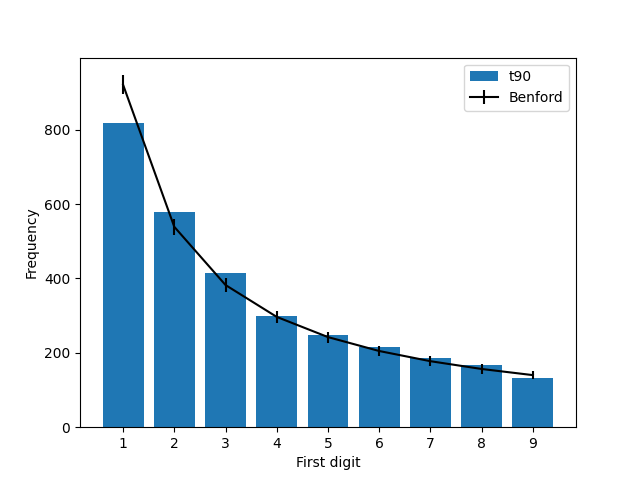}
  \includegraphics[width=70mm]{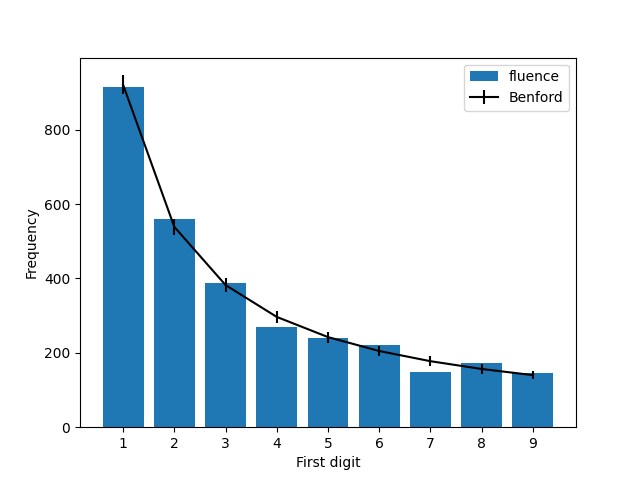}
  \vskip-0.1in
\caption{Same as Figure~\ref{fig1}, but now for $3061$ long GRBs (with $T_{90}>2$ s)
from the Fermi-GBM burst catalog.}
\label{fig2}
\end{figure*}

\begin{figure*}
\vskip-0.2in
   \centering
  \includegraphics[width=70mm]{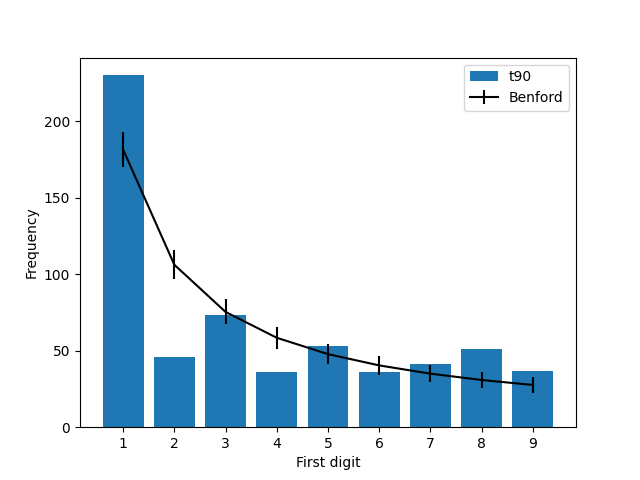}
  \includegraphics[width=70mm]{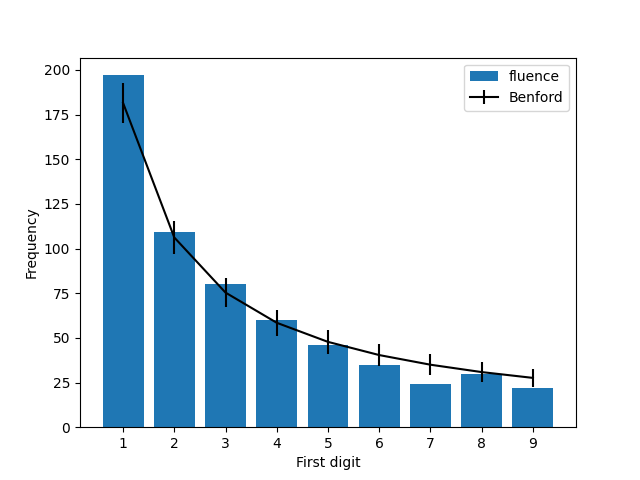}
  \vskip-0.1in
\caption{Same as Figure~\ref{fig1}, but now for $603$ short GRBs (with $T_{90}\leq2$ s)
from the Fermi-GBM burst catalog.}
\label{fig3}
\end{figure*}

GRBs can be divided into two classes, long GRBs and short GRBs, with a division line at $T_{90}=2$ s \citep{1993ApJ...413L.101K}.
Using the conventional division between the long and short GRB groups ($T_{90}>2$ s and $T_{90}\leq2$ s, respectively),
we find that there were $3061$ long GRBs and $603$ short GRBs in the Fermi-GBM burst catalog. In this work, we also check
if the durations and fluences of long and short GRBs obey Benford's law. The first digit distributions of the duration
and fluence of long and short GRBs are depicted in Figures~\ref{fig2} and \ref{fig3}, respectively. As listed in Table~\ref{tab1},
the $\chi^2$ tests and the corresponding $p$-values for the fluences are extremely supportive to the null hypothesis that
the fluences of both long and short GRBs conform to Benford's law. However, the durations of both long and short GRBs
obviously deviate from Benford's law. We find that the Pearson $\chi^2$ for the durations of $3061$ long GRBs and $603$
short GRBs are $19.9$ and $74.2$, which correspond to $p$-values of 0.01 and $7\times10^{-13}$, respectively.
On the basis of the $p$-values, we can safely reject the null hypothesis.

\begin{figure}
\vskip-0.2in
   \centering
  \includegraphics[width=70mm]{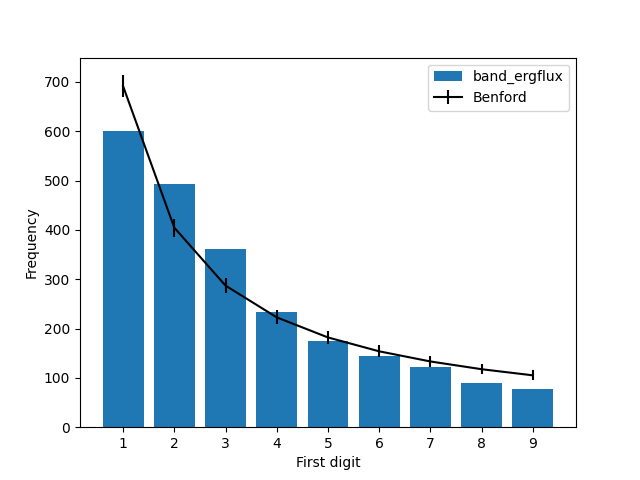}
  \vskip-0.1in
\caption{Distribution of the first digit of the energy flux of $2298$ available GRBs from the Fermi-GBM burst catalog.
The theoretical prediction from Benford's law (solid line) along with associated binomial errors are also shown for comparisons.}
\label{fig4}
\end{figure}

Similarly, in Figure~\ref{fig4} and Table~\ref{tab1}, we show the first digit distribution of the energy flux of
$2298$ available GRBs. As we can see that the relative rank of the probability of occurrence of leading digits
roughly agrees with Benford's law, but the Benford distribution is not scrupulously obeyed. The rather large
$\chi^2$ value and the extremely low $p$ value suggest that the energy flux of GRBs does not adhere to Benford's law.

\begin{table}
\begin{center}
\caption[]{Summary of our Benford analyses on the first digit distributions of the duration, fluence, and energy flux of GRBs observed with Fermi-GBM.}
\label{tab1}
  \begin{tabular}{lcccc}
  \hline\noalign{\smallskip}
   Dataset  &  Physical quantity  &   Number  &  $\chi^{2}/dof$  &  $p$-value  \\
  \hline\noalign{\smallskip}
  \multirow{2}{*}{All GRBs} & Duration  & 3364 & 13.0/8  &  0.11\\
                            & Fluence   & 3364 & 12.5/8  &  0.13\\
  \hline
  \multirow{2}{*}{Long GRBs} & Duration  & 3061 & 19.9/8  &  0.01\\
                             & Fluence   & 3061 & 11.5/8  &  0.18\\
  \hline
  \multirow{2}{*}{Short GRBs} & Duration  & 603 & 74.2/8  &  $7\times10^{-13}$\\
                              & Fluence   & 603 & 7.1/8  &  0.53\\
  \hline
  All GRBs & Energy flux  & 2298 & 67.4/8  &  $2\times10^{-11}$\\
  \noalign{\smallskip}\hline
  \end{tabular}
\end{center}
\end{table}

\section{Summary}
\label{sec:summary}
In this work, we have performed a systematic investigation on the first digit distributions of the duration, fluence,
and energy flux of GRBs. For the complete GRB sample detected by Fermi-GBM, our results show that the first digits of the duration
and fluence are not uniformly distributed, but that small digits are more common than big ones
according to a logarithmic distribution as expected by Benford's law. However, not all quantities obey
Benford's law. Artificial and restricted quantities often deviate from the law, such as the energy fluxes of
GRBs in our study. The main reason for the deviation may be that the energy flux measurements are restricted
by spectral fits for each GRB, and only 63\% of bursts have the required spectral information. Thus, there is
not enough dynamic range to make the energy fluxes be fully compliant with Benford's law.

The Fermi-GBM burst catalog also revealed a bimodal $T_{90}$ distribution, and two classes of GRBs, i.e.,
long versus short GRBs with a separation at about 2 s, were confirmed \citep{1993ApJ...413L.101K,2020ApJ...893...46V}.
Using the conventional long--short GRB classification scheme, we found that there were $3061$ long GRBs
(with $T_{90}>2$ s) and $603$ short GRBs (with $T_{90}\leq2$ s).
Here, the first digits of the duration and fluence of long and short GRBs were also examined for adherence to
Benford's distribution. We found that the fluences of both long and short GRBs still follow Benford's law,
but their durations are no longer consistent with this law. Our results indicate that the data of fluence
seems to be very natural and believable, but that $T_{90}$ is not always a good quantity to conduct GRB
classification. That is, the long--short GRB classification scheme does not directly represent the
intrinsic physical classification scheme \citep{2007ApJ...655L..25Z,2010ApJ...725.1965L,2013ApJ...763...15Q}.

To cross-check of the results produced on Fermi-GBM, we also analyzed the first digit distributions
for the GRB data observed with the Burst Alert Telescope (BAT) onboard the Swift satellite
(please see Appendix~\ref{Swift} for more details). We found that (i)
the derived $p$-values change quantitatively, though the qualitative results and conclusions remain the same for
the duration and fluence distributions of the overall GRB data, independent of what kind of GRB mission is considered;
(ii) Benford's law is still followed by the fluence distributions of both long and short GRBs, for each GRB mission;
(iii) the duration distributions of the long and short GRB groups observed with Swift-BAT are generally consistent with
Benford's law, but those of the Fermi-GBM sample deviate from this law. Since the duration distributions of long
and short GRBs for samples observed with different missions do not always apply to Benford's law, we emphasize agian
that the duration classification scheme does not always match the intrinsic physical classification scheme.


\normalem
\begin{acknowledgements}
We are grateful to the anonymous referee for helpful comments.
This work is partially supported by the Strategic Priority Research Program of the Chinese Academy
of Sciences (grant No. XDB0550400), the National Natural Science Foundation of China (grant Nos. 12373053
and 12321003), the Key Research Program of Frontier Sciences (grant No. ZDBS-LY-7014)
of Chinese Academy of Sciences, and the Natural Science Foundation of Jiangsu Province (grant No. BK20221562).
\end{acknowledgements}

\bibliographystyle{raa}
\bibliography{bibtex}



\appendix

\section{First digit distributions for Swift-BAT GRBs}
\label{Swift}

To cross-check the results produced on Fermi-GBM, we also analyze the first digit distributions
for the GRB data observed with Swift-BAT. For the Swift-BAT sample, the durations (in units of s) and fluences
(in units of erg $\rm cm^{-2}$) in the 15--150 keV energy range are taken from the online burst
catalog.\footnote{\url{https://swift.gsfc.nasa.gov/archive/grb_table/}} The dataset contains 1627 GRBs up to
2024 January. We remove 143 bursts for which no duration or fluence measurements were available. In total,
there are 1484 GRBs (including 1354 long bursts with $T_{90}>2$ s and 130 short ones with $T_{90}\leq2$ s)
for us to perform the first digit analyses.

\begin{figure*}
\vskip-0.2in
   \centering
  \includegraphics[width=70mm]{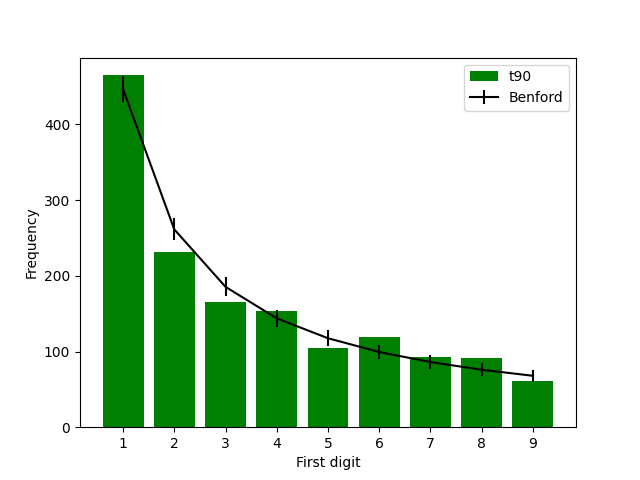}
  \includegraphics[width=70mm]{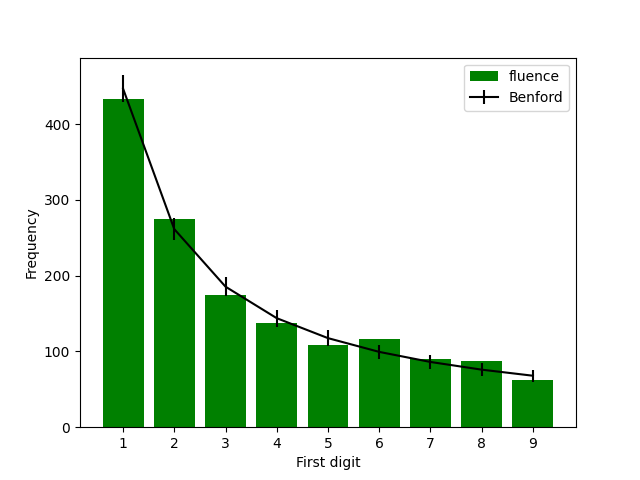}
  \vskip-0.1in
\caption{Distributions of the first digit of the duration $T_{90}$ and fluence of all $1484$ GRBs
from the Swift-BAT burst catalog. The theoretical predictions from Benford's law (solid lines) along with associated
binomial errors are also shown for comparisons.}
\label{figA1}
\end{figure*}

\begin{figure*}
\vskip-0.2in
   \centering
  \includegraphics[width=70mm]{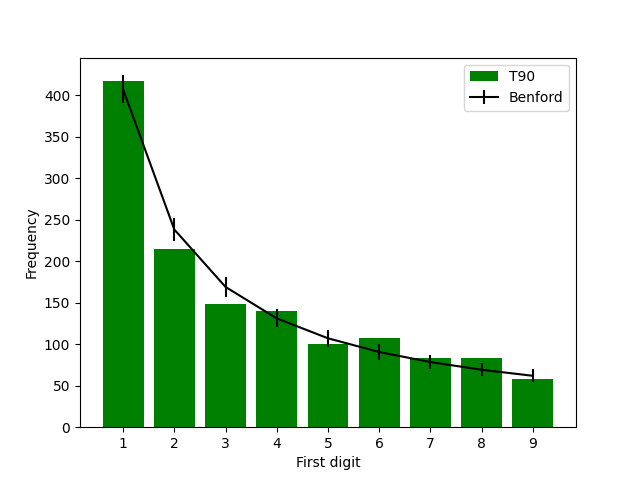}
  \includegraphics[width=70mm]{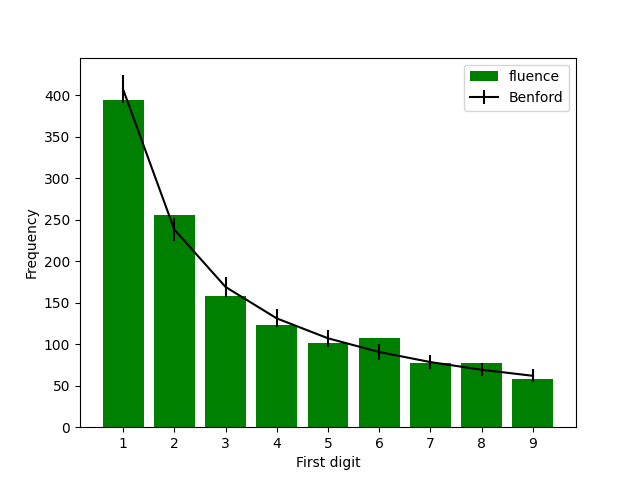}
  \vskip-0.1in
\caption{Same as Figure~\ref{figA1}, but now for $1354$ long GRBs (with $T_{90}>2$ s) from the Swift-BAT burst catalog.}
\label{figA2}
\end{figure*}

\begin{figure*}
\vskip-0.2in
   \centering
  \includegraphics[width=70mm]{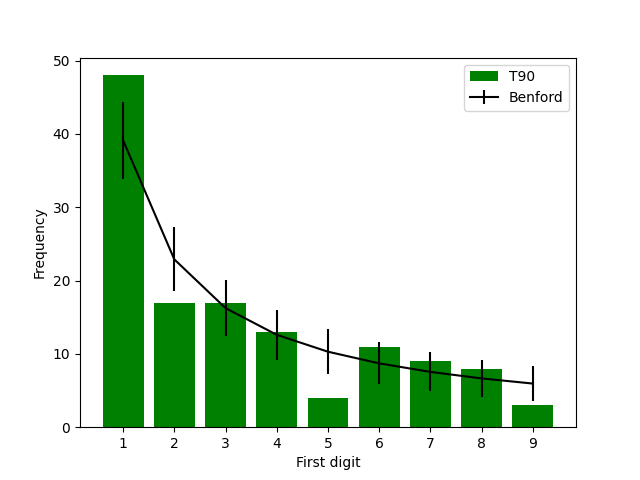}
  \includegraphics[width=70mm]{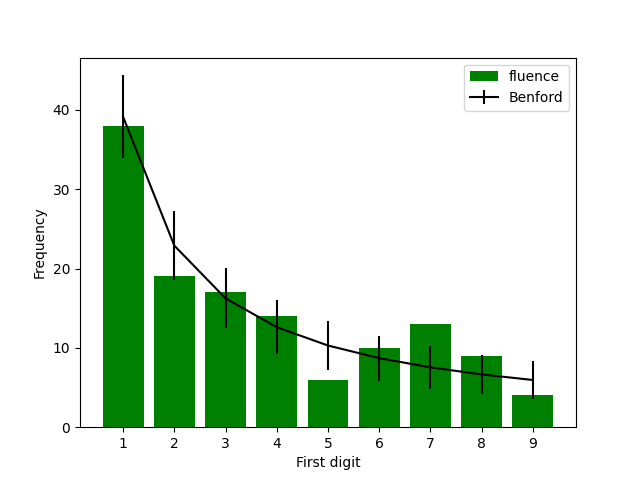}
  \vskip-0.1in
\caption{Same as Figure~\ref{figA1}, but now for $130$ short GRBs (with $T_{90}\leq2$ s) from the Swift-BAT burst catalog.}
\label{figA3}
\end{figure*}

The first digit distributions of the duration and fluence for all 1484 GRBs, 1354 long GRBs, and
130 short GRBs are illustrated in Figures~\ref{figA1}, \ref{figA2}, and \ref{figA3}, respectively. A tabular summary
of our Benford analyses for the Swift-BAT sample can be found in Table~\ref{tabA1}. The Pearson $\chi^2$ tests and
the corresponding $p$-values indicate that the durations and fluences for the overall sample and the subsamples of
long and short GRBs are all roughly consistent with Benford's law. As shown in Figure~\ref{figA3}, the digits 2, 5,
and 9 are smaller than the expected Benford distributions. That is, the duration and fluence of short GRBs do not
seem to fit Benford's law through the eyes, but their corresponding $p$-values support they do. The optical illusion
may be caused by the relatively small sample size of short GRBs.

The comparison between the Fermi-GBM and Swift-BAT samples may be summarized as follows:
(i) the first digit distributions of the duration and fluence of the overall GRB data conform to Benford's law,
independent of what kind of GRB mission is considered; (ii) Benford's law is still followed by the fluence
distributions of both long and short GRBs, for each GRB mission; (iii) the duration distributions
of the long and short GRB groups observed with Fermi-GBM do not obey Benford's law, but those of the Swift-BAT
sample are generally consistent with this law. Note that the ratios of short-to-long GRB numbers with a division
of $T_{90}=2$ s for the Fermi-GBM and Swift-BAT samples are 603:3061 (1:5.1) and 130:1354 (1:10.4), respectively.
Obviously, the $T_{90}$ distribution is instrument dependent. Again, our results suggest that the duration classification
scheme does not always match the intrinsic physical classification scheme. It it does, then the duration distributions
of both long and short GRBs for samples observed with different missions should always adhere to Benford's law.

\begin{table}
\begin{center}
\caption[]{Summary of our Benford analyses on the first digit distributions of the duration and fluence of GRBs detected by Swift-BAT.}
\label{tabA1}
  \begin{tabular}{lcccc}
  \hline\noalign{\smallskip}
   Dataset  &  Physical quantity  &   Number  &  $\chi^{2}/dof$  &  $p$-value  \\
  \hline\noalign{\smallskip}
  \multirow{2}{*}{All GRBs} & Duration  & 1484 & 16.4/8  &  0.04\\
                            & Fluence   & 1484 & 8.3/8  &  0.41\\
  \hline
  \multirow{2}{*}{Long GRBs} & Duration  & 1354 & 12.7/8  &  0.12\\
                             & Fluence   & 1354 & 7.5/8  &  0.48\\
  \hline
  \multirow{2}{*}{Short GRBs} & Duration  & 130 & 10.0/8  &  0.26\\
                              & Fluence   & 130 & 8.3/8  &  0.41\\
  \noalign{\smallskip}\hline
  \end{tabular}
\end{center}
\end{table}

\end{document}